\documentclass[iop]{emulateapj}

\usepackage{graphicx}
\usepackage{epstopdf}

\usepackage{amssymb}

\usepackage{amsmath}
\usepackage{amssymb}

\def\micron{$\mu$m}

\begin{document}

\title{ARCONS: A 2024 Pixel Optical through Near-IR Cryogenic Imaging Spectrophotometer}

\author{B.A. Mazin, S.R. Meeker, M.J. Strader, P. Szypryt, D. Marsden, J.C. van Eyken, G.E. Duggan, G. Ulbricht, and M. Johnson}
\affil{Department of Physics, University of California, Santa Barbara, CA 93106, USA}
\email{bmazin@physics.ucsb.edu}

\author{B. Bumble}
\affil{NASA Jet Propulsion Laboratory, 4800 Oak Grove Drive, Pasadena, CA 91125, USA}

\author{K. O'Brien}
\affil{Department of Physics, University of Oxford, Denys Wilkinson Building, Keble Road, Oxford, OX1 3RH, UK}

\author{C. Stoughton}
\affil{Fermilab Center for Particle Astrophysics, Batavia, IL 60510, USA}


\begin{abstract}

We present the design, construction, and commissioning results of ARCONS, the \textbf{Ar}ray \textbf{C}amera for \textbf{O}ptical to \textbf{N}ear-IR \textbf{S}pectrophotometry.  ARCONS is the first ground-based instrument in the optical through near-IR wavelength range based on  Microwave Kinetic Inductance Detectors (MKIDs). MKIDs are revolutionary cryogenic detectors, capable of detecting single photons and measuring their energy without filters or gratings, similar to an X-ray microcalorimeter.  MKIDs are nearly ideal, noiseless photon detectors, as they do not suffer from read noise or dark current and have nearly perfect cosmic ray rejection.  ARCONS is an Integral Field Spectrograph (IFS) containing a lens-coupled 2024 pixel MKID array yielding a 20"${\times}$20" field of view, and has been deployed on the Palomar 200" and Lick 120" telescopes for 24 nights of observing.  We present initial results showing that ARCONS and its MKID arrays are now a fully operational and powerful tool for astronomical observations. 
	
\end{abstract}


\section{Introduction}
\label{sec:intro}

Detectors and instrumentation have always been the biggest drivers of sensitivity improvements in astrophysics.  While the collecting area of ground-based telescopes has only improved by a factor of 4 in the last 50 years, the per-pixel sensitivity of astronomical detectors has grown by at least a factor of 20, and the improvements in quality and size have been even more impressive.  These improvements have led to a golden age in astrophysics, but we are rapidly reaching a plateau in the per-pixel performance of traditional semiconductor based (CCD \citep{Smith:2011hu} and HgCdTe \citep{2012SPIE.8453E..0WH}) detectors, with improvements only coming in the total size and pixel count of the final mosaic \citep{2009arXiv0912.0201L,2012SPIE.8446E..11F}.

To maintain the rapid pace of advance in per-pixel performance of detectors for astrophysics requires breaking away from semiconductors-based detectors, and developing technologies with potentially greater performance; for instance by reducing read noise and dark current, having a wider bandwidth, or having inherent spectral resolution.  The most promising avenue is to use superconducting detectors~\citep{Moseley:1988bk,Irwin:1996vk,Li:2001dc,2003Natur.425..817D} These detectors significantly increase performance by operating below 4 Kelvin, thereby reducing the contribution from thermal noise, allowing them to measure the energy and arrival time of a single photon with no false counts.  

The \textbf{Ar}ray \textbf{C}amera for \textbf{O}ptical to \textbf{N}ear-IR \textbf{S}pectrophotometry (ARCONS) was built around a highly multiplexible type of Low Temperature Detector (LTD) known as Microwave Kinetic Inductance Detectors (MKIDs)~\citep{2003Natur.425..817D,Mazin:2012kl}, optimized for optical and near-infrared astronomy.   The goal of ARCONS is to demonstrate the viability of both the MKID technology as well as do science that would be difficult or impossible with conventional instruments.   

An MKID array gives ARCONS significant advantages over a conventional lenslet, fiber fed, or image slicer integral field spectrograph (IFS) such as:

\begin{itemize}
\setlength{\itemsep}{1pt}
\item Simple optical design that enables very high throughput
\item Time resolution up to six orders of magnitude better than a CCD
\item Extremely broad intrinsic bandwidth (0.1--5 $\mu$m) with good quantum efficiency
\item No read noise or dark current, and nearly perfect cosmic ray rejection
\item No observing time is lost to readout of the array
\item Simpler scaling to much larger arrays than conventional IFSs.  For example, HARMONI~\citep{2012SPIE.8450E..1NT} for the E-ELT will use 8 4k${\times}$4k detectors but only has 32768 spaxels.  In an MKID, each pixel is a spaxel, albeit at significantly lower spectral resolution than an IFS like HARMONI. 
\item Time domain information allows \emph{ex post facto} use of calibration stars for monitoring atmospheric transparency, setting dynamic apertures, and applying tip/tilt corrections.  This improves signal to noise and calibration accuracy, and allows the instrument to make optimal use of the atmospheric conditions.
\item Photon arrival time, spectral resolution, and the large number of pixels allow for monitoring and the removal of sky emission, potentially down to the Poisson limit, even in spectral regimes dominated by time variable OH emission 
\end{itemize}

The MKID technology in ARCONS is relatively immature, leaving significant room to improve future instruments.  Current optical MKID technology has a spectral resolution R=$\lambda/ \Delta \lambda {\sim}10$ at 4000 \AA.  Improvements in the MKID and readout electronics are on track to increase the spectral resolution towards the theoretical limit of around 100 for a 100 mK operating temperature.  The array size, though small compared to an imaging CCD (although not small compared to IFSs), is currently only limited by what we can afford to readout with our room temperature electronics (Section~\ref{sec:readout}).  As the components used in these electronics are improving with Moore's Law, megapixel arrays should be possible within a decade.

We describe the MKID arrays in detail in Section~\ref{sec:MKID}.  The optical design is presented in Section~\ref{sec:optical}, and the instrument design, including cryogenics, mounting hardware, guide camera, and wavelength calibration system is presented in Section~\ref{sec:inst}.  Section~\ref{sec:readout} describes the cryogenic and room temperature readout electronics, and Section~\ref{sec:pipeline} contains a brief outline of the data storage and processing.  Section~\ref{sec:perf} contains on-sky performance results. 

\section{Microwave Kinetic Inductance Detectors}
\label{sec:MKID}

\begin{figure*}
\begin{center}
\includegraphics[width=1.6\columnwidth]{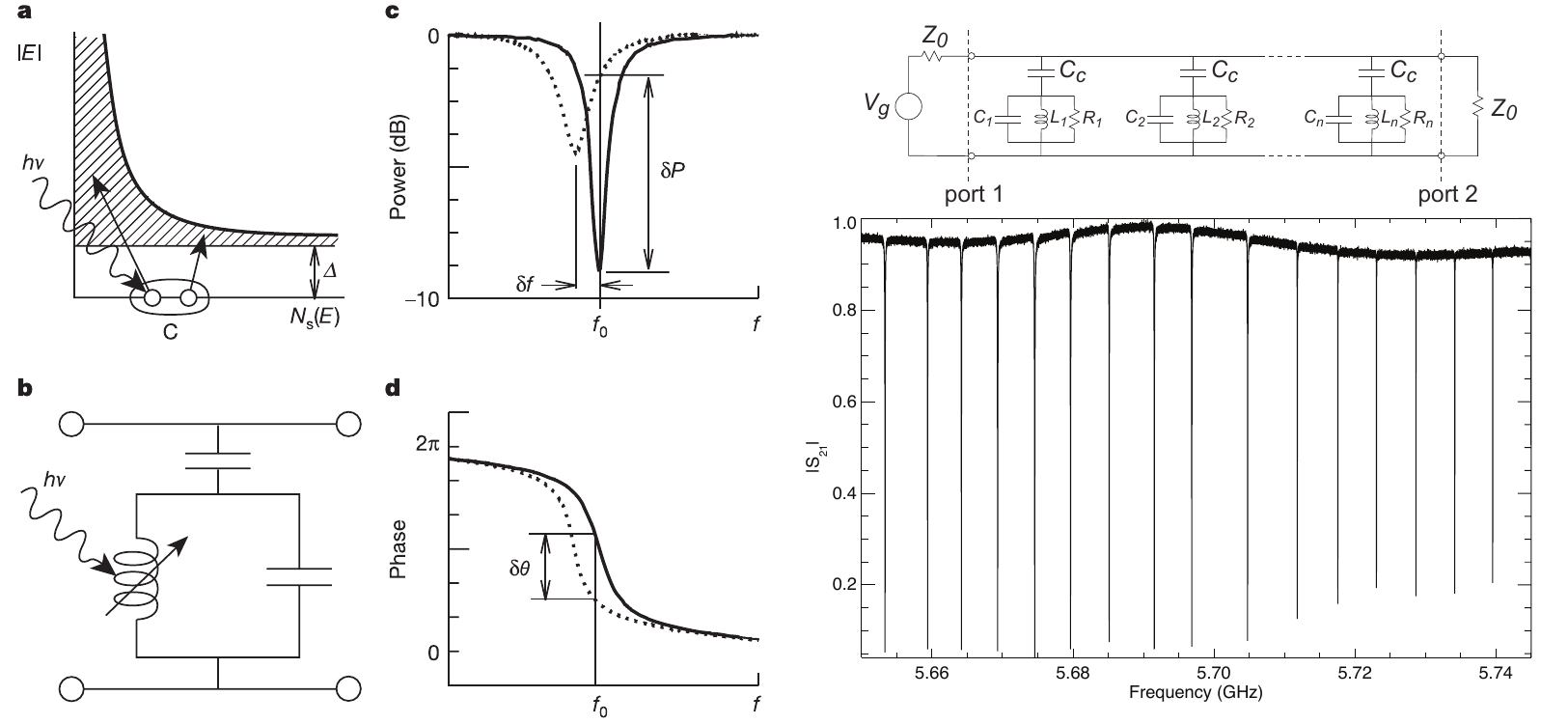}
\end{center}
\caption{Left: The basic operation of an MKID, reprinted from \citep{2003Natur.425..817D}. (a) An energy level diagram showing the number of states $N_s(E)$ of a superconductor with gap $\Delta$. Photons with energy $h\nu$ are absorbed in a superconducting film, breaking Cooper pairs $C$, and producing a number of excitations called quasiparticles.  (b) To sensitively measure these quasiparticles, the film is placed in a high frequency planar resonant circuit.  In right panels the amplitude (c) and phase (d) response of a microwave excitation signal sent through the resonator as a function of frequency is shown.  The change in the surface impedance of the film following a photon absorption event pushes the resonance to lower frequency and changes its amplitude.  If the detector (resonator) is excited with a constant on-resonance microwave signal, the energy of the absorbed photon can be determined by measuring the degree of phase and amplitude shift.  Right: An example of frequency domain multiplexing (FDM) of MKIDs showing many resonators being read out through a single transmission line.} \label{fig:detcartoon}
\end{figure*}

Low temperature detectors (LTDs) with operating temperatures on the order of 100~mK, are currently the preferred technology for astronomical observations over most of the electromagnetic spectrum, notably in the far infrared through millimeter (0.1--3~mm)~\citep{Bintley:2010gz,Niemack:2008gk,Carlstrom:2011ik}, X-ray~\citep{Kelley:2009ce}, and gamma-ray~\citep{Doriese:2007hya} wavelength ranges.  In the important Ultraviolet, Optical, and Near-Infrared (UVOIR) (0.1--5~$\mu$m) wavelength range a variety of detector technologies based on semiconductors, backed by large investment from both consumer and military customers, has resulted in detectors for astronomy with large formats, high quantum efficiency, and low readout noise.  These detectors, however, are fundamentally limited by the large band gap of the semiconductor which restrict the maximum detectable wavelength (1.1 eV for silicon) and thermal noise sources from their relatively high ($\sim$100~K) operating temperatures.  LTDs allow the use of superconductors with gap parameters over 1000 times lower than semiconductors.  This difference allows a leap in capabilities.  A superconducting detector can count single photons with no false counts while determining the energy (to several percent or better) and arrival time (to a microsecond) of the photon (the optical analog of an X-ray calorimeter).  It can also have much broader wavelength coverage since the photon energy is always much greater than the gap energy.  While a CCD is limited to about 0.3--1~$\mu$m, the MKIDs described here are in principle sensitive from 0.1~$\mu$m in the UV to greater than 5~$\mu$m, enabling observations at important infrared wavelengths.  

Superconducting UVOIR detectors have been pursued in the past with two technologies, Superconducting Tunnel Junctions (STJs)~\citep{Martin:2006dx,Hijmering:2008kn} and Transition Edge Sensors (TESs)~\citep{Romani:2001fw,Burney:2006ds}.  While both of these technologies result in functional detectors, they are limited to single pixels or small arrays due to the lack of a credible strategy for wiring and multiplexing large numbers of detectors, although recently there have been proposals for larger TES multiplexers~\citep{Niemack:2010gs}.
  
Microwave Kinetic Inductance Detectors, or MKIDs \citep{2003Natur.425..817D,Mazin:2012kl}, are a newer cryogenic detector technology that has proven important for astrophysics~\citep{Schlaerth:2010ex,2010A&A...521A..29M} due to their sensitivity and the ease with which they can be multiplexed into large arrays.  The ``microwave'' in MKIDs comes from their use of frequency domain multiplexing~\citep{Mazin:2006jy} at microwave frequencies (0.1--20 GHz) that allows thousands of pixels to be read out over a single microwave cable. The Optical Lumped Element (OLE)~\citep{Doyle:2008gc} MKID arrays we have developed have significant advantages over semiconductor detectors.  They can count individual photons with no false counts and determine the energy and arrival time of every photon with good quantum efficiency.  Their physical pixel size and maximum count rate is well matched with large telescopes.  These capabilities enable powerful new astrophysical instruments usable from the ground and space. 

MKIDs work on the principle that incident photons change the surface impedance of a superconductor through the kinetic inductance effect~\citep{mattis58}.  The kinetic inductance effect occurs because energy can be stored in the supercurrent (the flow of Cooper Pairs) of a superconductor.  Reversing the direction of the supercurrent requires extracting the kinetic energy stored in it, which yields an extra inductance term in addition to the familiar geometric inductance.  The magnitude of the change in surface impedance depends on the number of Cooper Pairs broken by incident photons, and is hence proportional to the amount of energy deposited in the superconductor. This change can be accurately measured by placing a superconducting inductor in a lithographed resonator.  A microwave probe signal is tuned to the resonant frequency of the resonator, and any photons which are absorbed in the inductor will imprint their signature as changes in phase and amplitude of this probe signal.  Since the quality factor $Q$ of the resonators is high and their microwave transmission off resonance is nearly perfect, multiplexing can be accomplished by tuning each pixel to a different resonant frequency with lithography during device fabrication.  A comb of probe signals can be sent into the device, and room temperature electronics can recover the changes in amplitude and phase without significant cross talk~\citep{2012RScI...83d4702M}, as shown in the right panel of Figure~\ref{fig:detcartoon}.

\begin{figure}
\begin{center}
\includegraphics[width=0.45\textwidth]{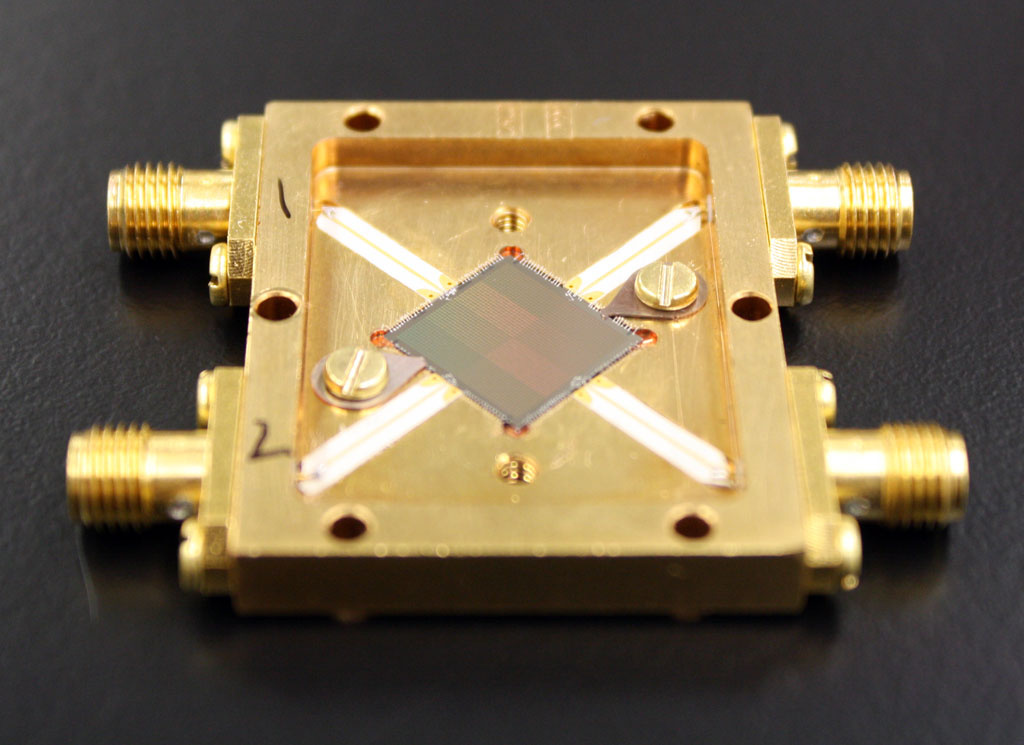}
\end{center}
\caption[MKID array.]{A photograph of a 2024 pixel UVOIR MKID array (with microlens removed for clarity) in a microwave package.}
\label{fig:mkids}
\end{figure}

The primary theoretical limitation on the spectral resolution is from the intrinsic quasiparticle creation statistics during the photon absorption event.  The energy from the photon can end up in two places, the quasiparticle system and the phonon system.  These systems interact, allowing energy exchange between the two, which reduces the statistical fluctuation from Poisson by the Fano factor $F$, typically assumed to be 0.2~\citep{fano47}. The spectral resolution of the detector, $R = \lambda/ \Delta \lambda =  E/\Delta E$, can be written as $R = \frac{1}{2.355}\sqrt{\frac{\eta h \nu}{F \Delta}}$, where $\eta$ is the efficiency of conversion of energy into quasiparticles, typically 0.57~\citep{Kozorezov:2007ek}, $h \nu$ is the photon energy, $F$ is the Fano factor, and $\Delta$ is the energy gap.  The energy gap depends on the superconducting transition temperature $(T_c)$ of the inductor, $\Delta \approx 1.72 k_B T_c$, and we typically operate at a base temperature of $T_c/8$.  Going to lower $T_c$, and hence lower operating temperature, improves the theoretical $R$.  Operating at 100 mK yields a theoretical spectral resolution of R${\approx}$100 at 400 nm.  Previous research with Superconducting Tunnel Junctions (STJs) with superconducting absorbers has shown that superconducting absorbers can approach the Fano limit~\citep{Li:2001dc,Huber:2004ga,leGrand:1998vh}.

MKIDs are very versatile, as essentially any resonator with a superconductor as the inductor will function as a MKID.  In 2008 we decided to pursue a lumped element resonator design~\citep{Doyle:2008gc}.  The resonator itself consists of a 60~nm thick~\citep{2012SPIE.8453E..0BM} sub-stoichiometric titanium nitride (TiN$_x$) film~\citep{2010ApPhL..97j2509L} with the nitrogen content tuned with $x<1$ such that the superconducting transition temperature $T_c$ is about 800 mK.  Due to the long penetration depth of these films ($\sim$1000 nm) the surface inductance is a high 25 pH/square, allowing a very compact resonator fitting in a 222$\times$222~$\mu$m square.  A square microlens array with a 92\% optical fill factor is used to focus light onto the photo-sensitive inductor.  The pixel pitch is easily controlled, with pitches between 75--500~\micron~relatively easy to achieve.  The quasiparticle lifetime in our TiN films is 50--100~$\mu$s.  This sets the pulse decay time, allowing a maximum count rate of approximately 2500 cts/pixel/second before problems arise in separating pulses.

After significant development~\citep{Mazin:2012kl}, we have arrived at the 2024 (44${\times}$46) pixel array shown in Figure~\ref{fig:mkids}.  Typical performance for single pixel devices is R${\sim}$10 at 4000 \AA.  Over an entire array, we see a median spectral resolution R${\sim}$8 at 4000 \AA~(Section~\ref{sec:res}).  Typically $>$90\% of the resonators show up in frequency sweeps, but due to the variations in the TiN gap there are a significant number of collisions (two or more pixels with resonant frequencies closer together than 500 kHz), reducing usable pixels, as discussed in Section~\ref{sec:yield}.  After cutting out collisions and pixels with especially high or low quality factor, we usually have ${\sim}$70\% of our pixels usable.  A more uniform film, such as the ones made at NIST with Ti/TiN multilayers~\citep{Vissers:2012wu} or with atomic layer deposition (ALD), could significantly reduce the number of collisions, dramatically improving yield.

Even with the current moderate spectral resolution and yield, these arrays are some of the most powerful tools for narrow field of view astrophysics ever developed. 

\begin{figure}
\begin{center}
\includegraphics[width=1.0\columnwidth]{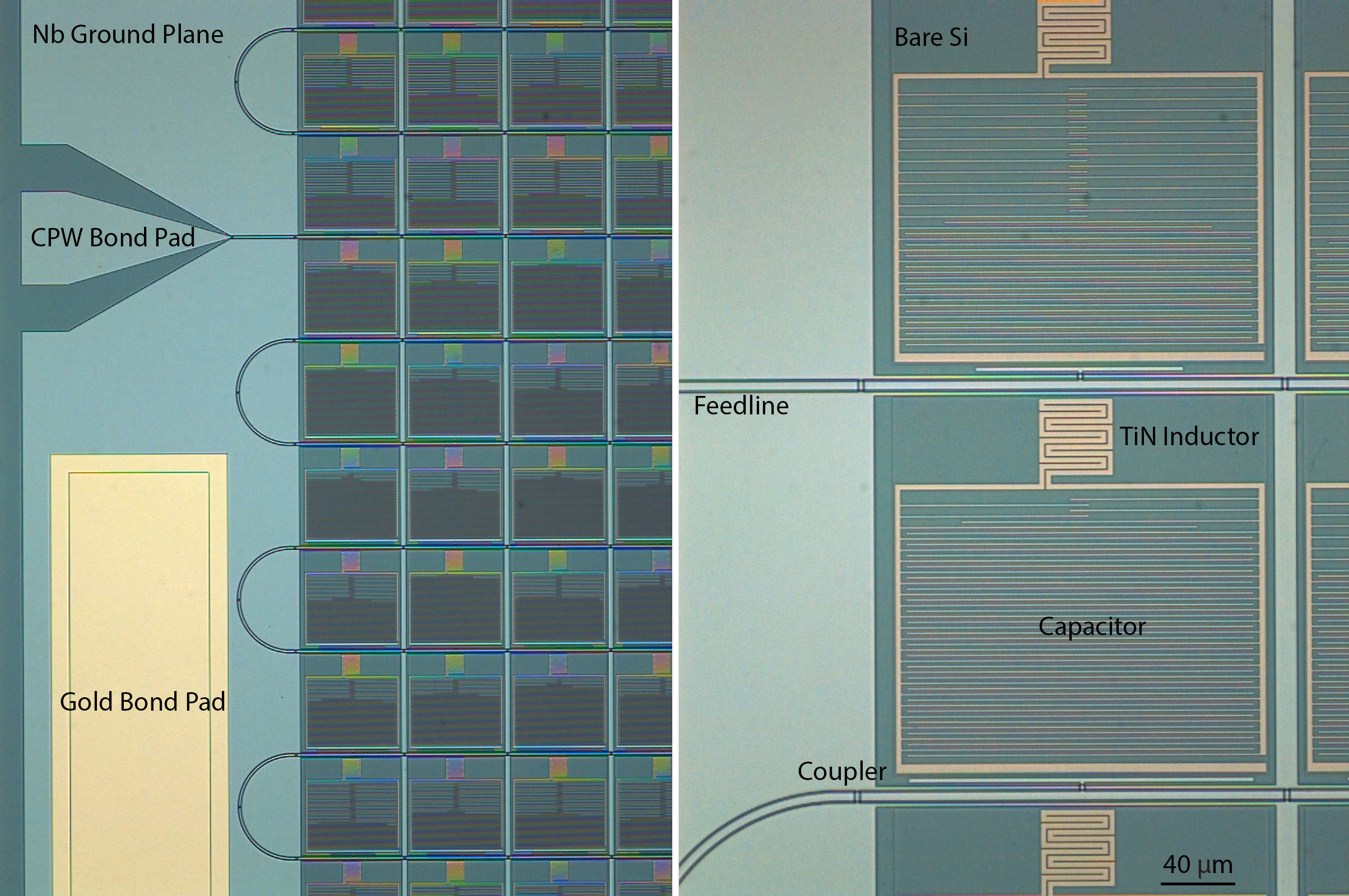}
\end{center}
\caption{Left: A microscope image of a portion of the 2024 pixel MKID array used at the Palomar 200" telescope. A microlens focuses the light on to the 40${\times}$40~\micron~inductor.  Various features of interest are labeled.  Right: A zoomed in view of the array in the left panel. } \label{fig:SCI4}

\end{figure}

\section{Optical Design}
\label{sec:optical}

\begin{figure}
\begin{center}
\includegraphics[width=1.0\columnwidth]{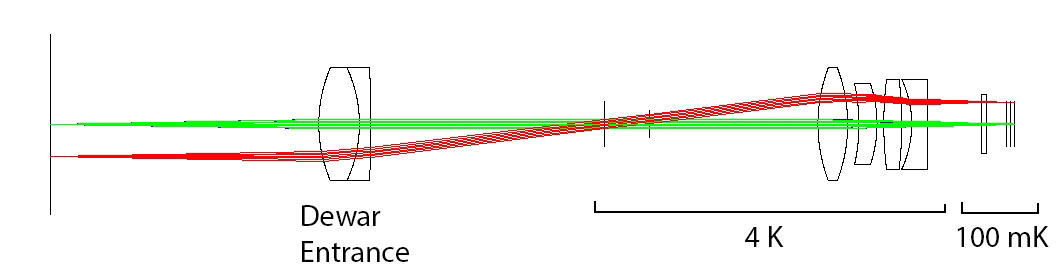}
\end{center}
\caption{The optical layout of ARCONS.  Light enters ARCONS at the left side of the figure, and is absorbed in the MKID array on the right side of the figure.}
\label{fig:raytrace}
\end{figure}

An optical system, shown in Figure~\ref{fig:raytrace}, has been designed with Zemax to couple ARCONS with the f/30 beam at the Palomar 200" Coud\'{e} focus.  A very similar system (using just a different dewar entrance lens) is used to couple to the Lick 3-m Coud\'{e} focus.  This focus is stationary, which makes interfacing with the cryostat (and its compressor) significantly easier, and the time resolution of the MKIDs means that rotation of the image plane over time can be easily removed.  

The optical path is as follows: light bounces off the primary and secondary, then off the Coud\'{e} M3 that directs light down the polar axis of the equatorial mount.  Once in the Coud\'{e} antechamber, a 100 mm diameter pickoff mirror (M4) sends the light towards the dewar entrance window.  Before the dewar entrance, two pickoff mirrors form a ${\sim}$25 arcsecond wide slit and direct the rest of the field of view towards the guide camera's reimaging optics.  Immediately in front of the entrance window is a commercial Thor Labs 1" filter wheel, which can hold up to 6 filters for calibration.  Observing with ARCONS is typically done at a filter wheel position that does not contain a filter.

The dewar entrance window is an achromatic doublet which collimates the light.  All the powered optics have a Edmund Optics VIS-NIR anti-reflection coating (reflection ${<}1$\% per surface between 4000-10000 \AA) except the BK7 meniscus lens (Surface 11 and 12 in Table~\ref{table:prescrip}) which is uncoated.   The light then passes through the dewar's 77 K shield though a baffle, and encounters the dewar's 4 K radiation shield.  At the 4 K radiation shield an Asahi SuperCold filter\footnote{http://www.asahi-spectra.com/opticalfilters\\/detail.asp?key=YSC1100} is placed to set the red edge of ARCONS's wavelength sensitivity.  This was set conservatively at 1.1~\micron~in order to reduce sky count rates, but in subsequent runs we will attempt to extend the wavelength coverage out through J band to 1.35~\micron.

There is a pupil image at 4 K suitable for placing a cold 2.1 mm diameter Lyot stop, but this turned out not to be necessary because the glass in the optics at 4 K blocks out enough of the 300 K thermal blackbody radiation to keep the 100 mK stage from warming up.  The SuperCold filter does effectively function as a cold Lyot stop with a diameter of 12 mm.  There are two achromatic doublets at 4 K to decollimate the beam and set the plate scale, but these achromats are physically separated to eliminate any potential issues with differential thermal contraction between the different glasses.  After the final 4 K optic, the light reaches the 100 mK stage.  Here the 4 K blackbody radiation is blocked out with a thin piece of borosilicate glass coated in Indium Tin Oxide (ITO), a transparent conductor.  The glass absorbs infrared up to roughly 400~\micron, and the ITO reflects a substantial portion of the longer wavelength radiation.  Finally, a square microlens with a 0.92 mm focal length and a 92\% optical fill factor focuses the light on the sensitive MKID inductor.  The optical prescription is shown in tabular form in Table~\ref{table:prescrip}. 

\begin{figure}
\begin{center}
\includegraphics[width=1.0\columnwidth]{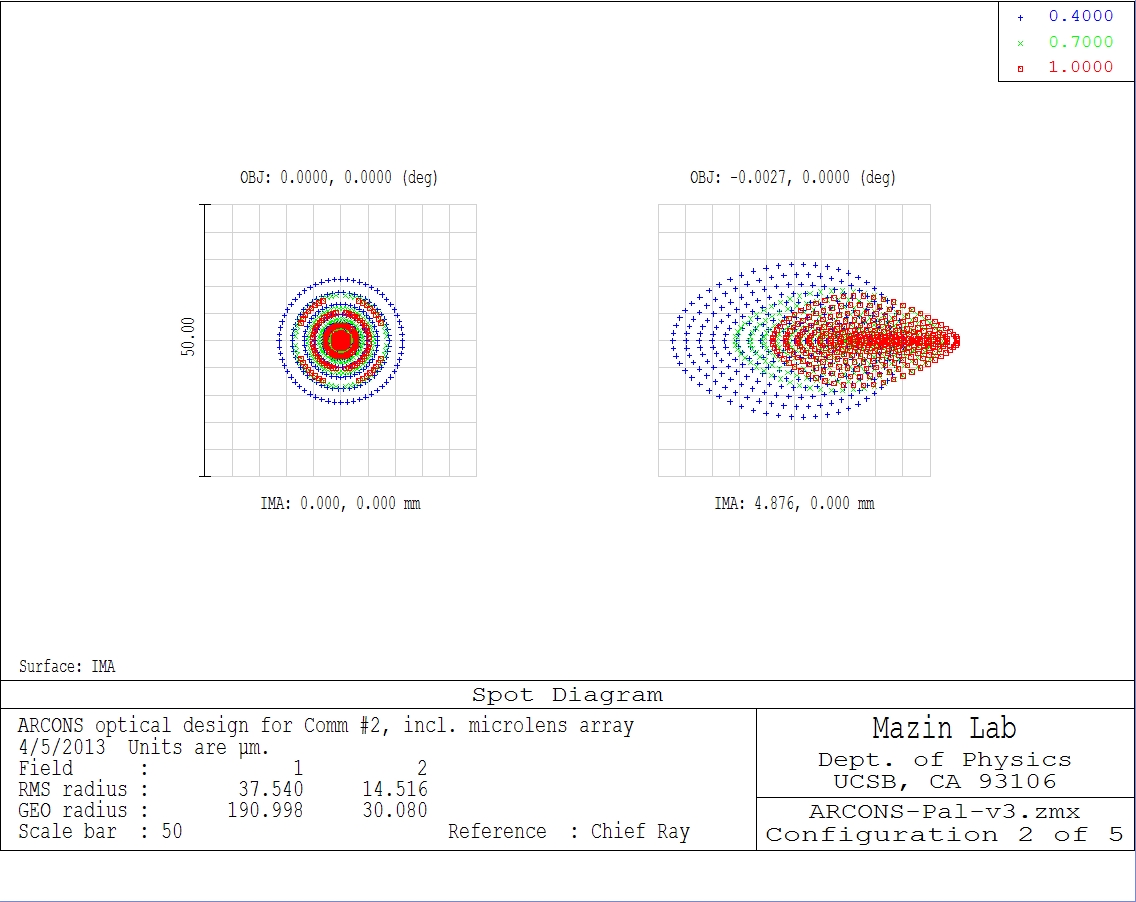}
\end{center}
\caption{Spot diagram of the ARCONS optics at 0.4 (blue), 0.7 (green), and 1.0~\micron~(red) plotted on grid of overall dimensions of 50${\times}$50~\micron.  The left panel shows the spot on axis, and the right panel shows the spot at the extreme right side of the array.  Even at the edges of the array the vast majority of the spot is contained within the 40${\times}$40~\micron~inductor.}
\label{fig:spot}
\end{figure}

The resulting spot diagram can be seen in Figure~\ref{fig:spot}.  The simulation includes the microlens array in front of the detector.  The MKIDs pixel pitch of 222~\micron~means that the optical system does not need high magnification, simplifying the system.  The system is designed to have 0.45" pixels on the sky.  The clear aperture of the optical system was designed to allow an unvignetted field of view of $20\times20$ arcseconds.  

This optical system was over-designed with every precaution possible to block out thermal blackbody radiation, which did not turn out to be required.  Future MKID cameras will likely completely eliminate the powered cold optics in favor of much simpler optical systems, including just warm optics to adjust the plate scale, a flat dewar window, a 4 K BK7 flat window, and a 100 mK ITO on BK7 IR blocker. 

\begin{table*}\footnotesize
\begin{center}
\begin{tabular}{lcccccl}
\hline
Surface & Type & Radius (mm) & Thickness (mm) & Glass & Diameter & Comment\\
\hline
1 & STANDARD & -46.44035 & -7 & N-BK7 & 25 & \\
2 & STANDARD & 33.77009 & -2.5 & N-SF5 & 25 & \\
3 & STANDARD & 95.94 &  -70.7 &  & 25 & \\
4 & STANDARD & Infinity & 2.538639 &  & 2.1 &  Lyot stop (not used) \\
5 & STANDARD & Infinity & -10 &  & 6  &  Supercold Filter\\
6 & STANDARD & Infinity & -37.3 &  & 6  &  \\
7 & STANDARD & -35.09 & -6.6 & N-BK7 & 25 &  \\
8 & STANDARD &  35.09 & -2.5 & & 25 & \\
9 & STANDARD &  41.18 & -4 & N-SF8 & 17 & \\
10 & STANDARD & 28.35 & -1.5 & & 18 & \\
11 & STANDARD & -82 & -4 & N-BK7 & 20 & \\
12 & STANDARD & 82 & -2.2 & & 20 & \\
13 & STANDARD & 23.54 & -3.5 & N-SF11 & 20 & \\
14 & STANDARD & Infinity & -12 & & 20 & \\
15 & STANDARD & Infinity & -1 & BK7 & 13 & 100 mK IR Blocker\\
16 & STANDARD & Infinity & -4.5 & & 13 & \\
17 & STANDARD & Infinity & -1 & LITHOSIL-Q & 10 & \\
18 & USERSURF & 0.43 & -0.78 & LENS ARRAY & 10 & APO-Q-P222-F0.93mm\\

\hline
\end{tabular}
\end{center}
\caption{The optical prescription for ARCONS.}
\label{table:prescrip}
\end{table*}

\section{Instrument Assembly and Mounting}
\label{sec:inst}

\subsection{Cryogenics}

ARCONS is built around an adiabatic demagnetization refrigerator (ADR) made by Janis Research.  The system contains a Cryomech PT-405 two stage pulse tube cooler.  The first stage of the pulse tube cools the outer radiation shield to 50--70 K, while the second stage cools the ADR to 3.3 K.  The pulse tube is powered by an air cooled compressor which can be located up to 100' from the cryostat.  At Palomar, the compressor is placed in a small room on the telescope floor that has a hatch that can be opened to vent the heat outside of the dome.  At Lick, the compressor is placed in the battery room outside the dome.

The ADR is a single shot magnetic cooler.  During the afternoon the superconducting ADR magnet is ramped to 40 kG, the heat switch is opened, and then the magnet is slowly ramped down during the course of the night.  A PID loop on a Lakeshore 370 controller provides input to a Lakeshore 625 Superconducting Magnet Power Supply, allowing us to stabilize the temperature of the ADR to better than 50~$\mu$K.  Stabilizing the temperature at 110 mK provides over 12 hours of hold time.  There are no obvious drifts in the MKIDs resonant frequencies as the magnetic field remaining in the superconducting magnet ramps down from 0.75 kG to 0 kG over the 12 hours hold time.

\subsection{Mounting to the Telescope}

\begin{figure}
\begin{center}
\includegraphics[width=1.0\columnwidth]{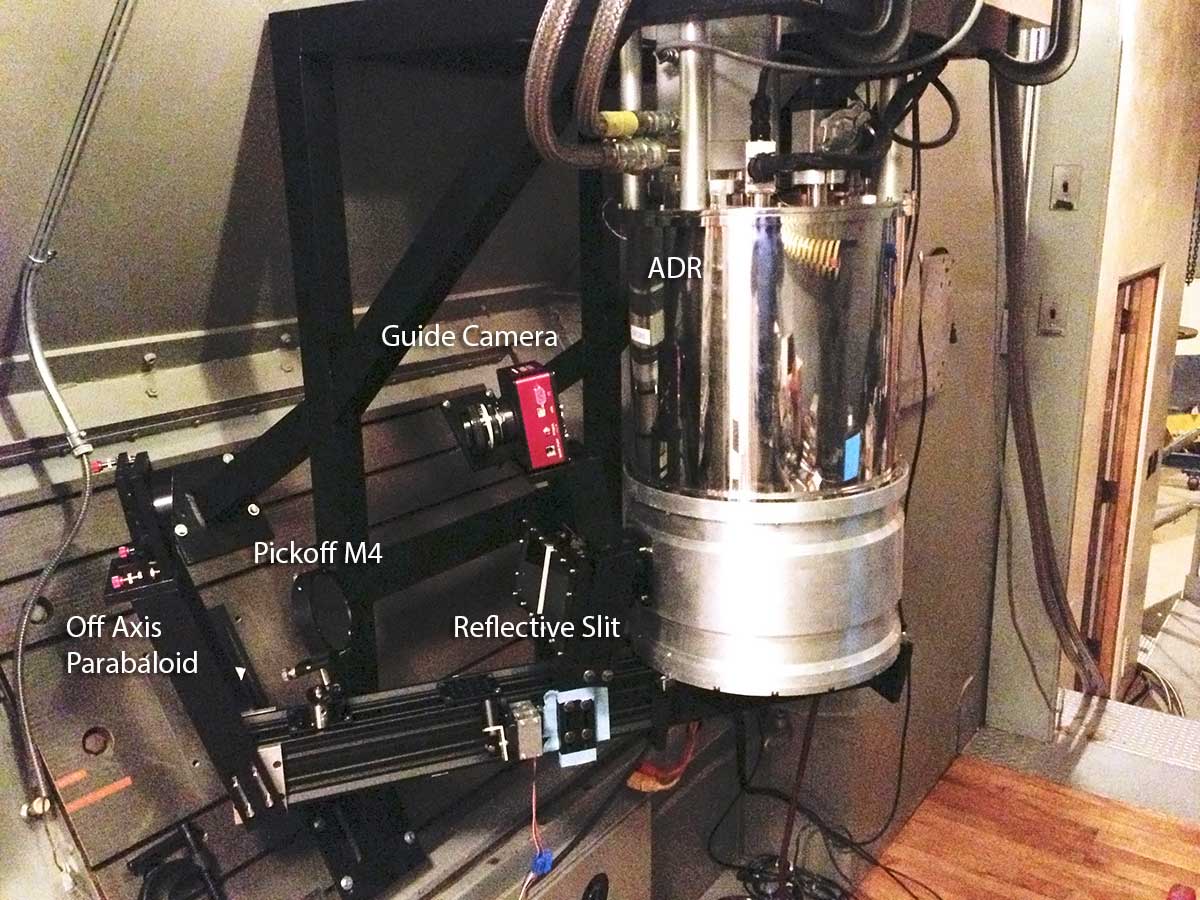}
\end{center}
\caption{ARCONS installed at the Palomar 200".}
\label{fig:arconsmount}
\end{figure}

At the Palomar 200", an aluminum frame is attached to a steel plate in the Coud\'{e} antechamber that is used to accommodate the Coud\'{e} spectrograph's slit jaws.  ARCONS hangs down from this aluminum frame, as shown in Figure~\ref{fig:arconsmount}.

At the Lick 3-m, ARCONS is also placed in the Coud\'{e} antechamber, but since the Hamilton Spectrograph is still in use the mounting plate is not used.  Instead, ARCONS is hung from an I-beam in the ceiling and then stabilized to the mounting plate with a steel strut.

\subsection{Guide Camera and Software}

The guide camera for ARCONS is a SBIG STF-8300M CCD Camera.  Light is reflected off mirrors on either side of the ARCONS entrance aperture, bounces to an off-axis parabaloid, and then goes through a Nikon 50 mm f/1.4 lens mounted to the SBIG camera.  The unvignetted field of view of the guide camera is about 1.5 arcminutes in diameter.  The SBIG camera, using 3${\times}$3 binning, is read out over USB by a small Windows computer located in the electronics rack.  In 10 second exposures we are able to guide off of stars in the 18--20$^{th}$ magnitude range.  For brighter objects, exposure times down to 0.1 seconds are used.  This computer is accessed by remote desktop in the control room.  The guiding software is a modified version of the Palomar Observatory's standard guiding package.

\subsection{Wavelength Calibration System}
\label{sec:wavecal}

As part of the data taking procedure each pixel's raw phase response must be mapped to photon energy.  We developed a system to uniformly illuminate the ARCONS focal plane with three lasers at the same time before and after science observations, thereby providing a reference to calibrate each detector's response.  

\begin{figure}
\begin{center}
\includegraphics[width=1.0\columnwidth]{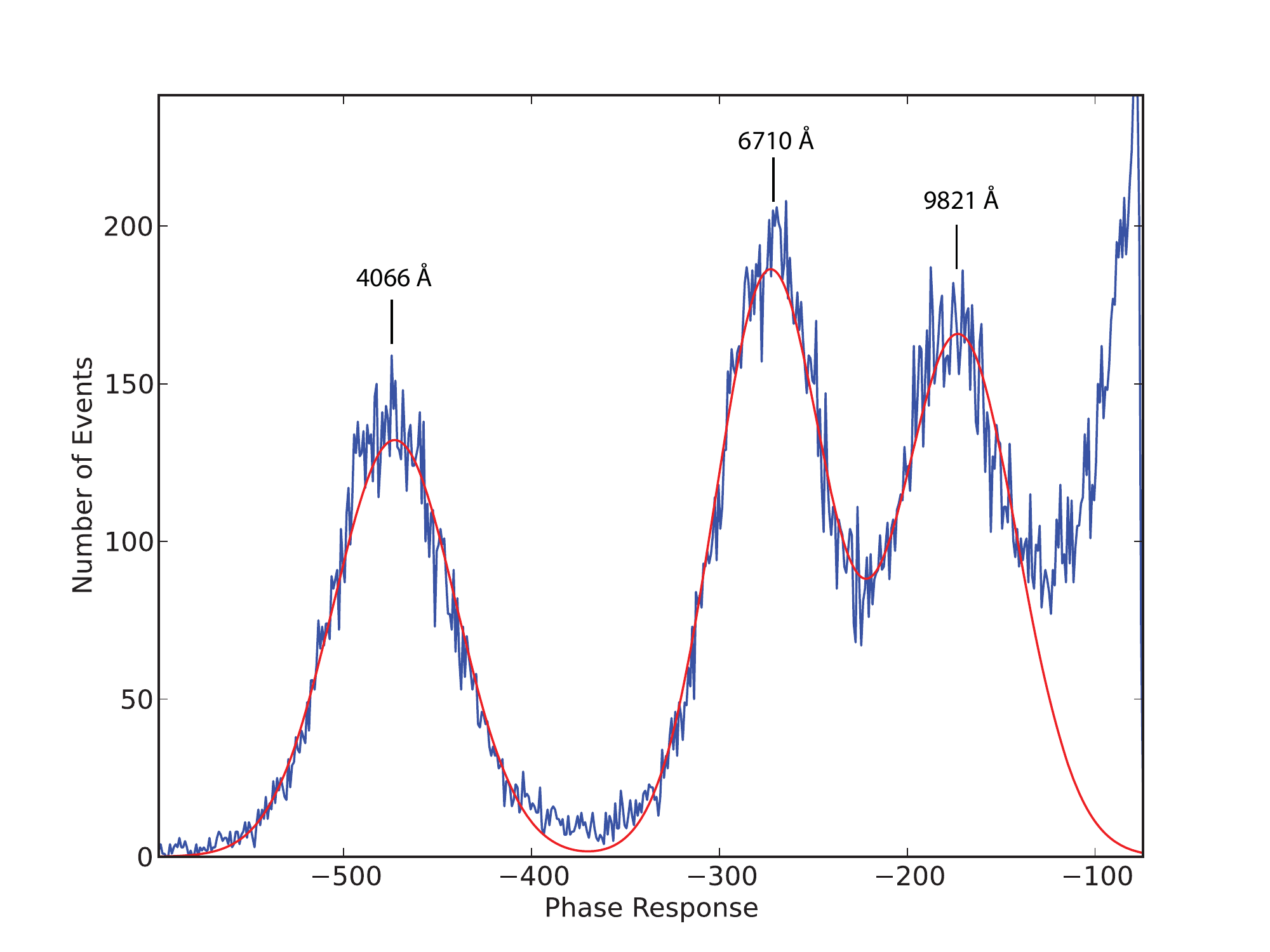}
\end{center}
\caption{A typical histogram of pulse heights returned by the wavelength calibration system is shown in blue, and a model fit to three gaussians is shown in red.  Lower values of phase response indicate higher energy photons.}
\label{fig:3peaks}
\end{figure}

The three lasers operate at 4066, 6710, and 9821~\AA.  They are integrated into a package that allows for adjustment of the intensity of each laser's light.  The light from the three lasers is combined in an integrating sphere with a fiber optic output.  The fiber optic carries the combined light to optics and diffusers that simulate the f/30 telescope beam.  A mirror on a mechanical arm can be rotated into position to block out the sky light and direct the calibration beam into the dewar.  The amount of exposure time and frequency of calibrations were controlled through software.  During observations, this process occurred approximately once every hour.

The response for a single detector is shown in Figure~\ref{fig:3peaks}, where the number of counts for a one minute integration are plotted against detector phase in blue, using bins of width one digital number as returned by the firmware.  Three Gaussians were simultaneously fit to the three peaks produced in response to the laser light, with the width of each Gaussian being a measure of the noise and therefore spectral resolution of the detector.  The fit is shown in red.  The tail out to low energy (less negative phase response) is due to the low energy background (thermal blackbody radiation, noise triggers, etc.) and is not included in the fit.  The location of the three peaks is then fit with a second order polynomial in phase and used to convert signals on that pixel to photon energy.  A given calibration solution is applied to observations near in time.  An analysis of the calibration solutions over time showed them to be consistent within 1 sigma between thermal cycles of the dewar or re-thresholding of the detectors.

\section{Readout}
\label{sec:readout}

\subsection{Room Temperature Readout}

\begin{figure}[h]
\begin{center}
\includegraphics[width=1.0\columnwidth]{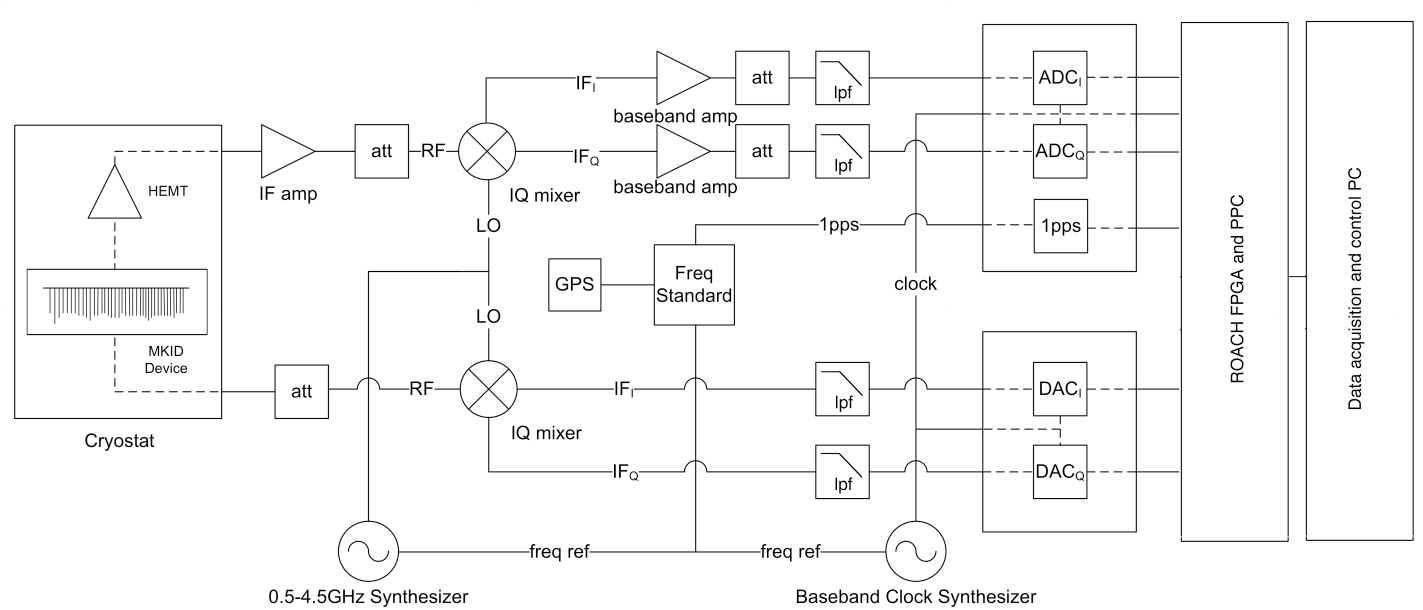}
\end{center}
\caption{A block diagram of the digital readout for ARCONS, reprinted from~\cite{2012RScI...83d4702M}. } 
\label{fig:readout}
\end{figure}

A block diagram for the digital readout for ARCONS is shown in Figure~\ref{fig:readout} and is described in detail in~\cite{2012RScI...83d4702M}. The digital readout for ARCONS utilizes eight Reconfigurable Open Architecture Computing Hardware (ROACH) boards produced by the Collaboration for Astronomy Signal Processing and Electronics Research (CASPER) \citep{2009astro2010T..21P}.  Each ROACH board houses a Virtex5 field programmable gate array (FPGA) and a PowerPC processor, and each board is connected to two digital-to-analog converters (DACs), two analog-to-digital converters (ADCs), and a circuit board used for mixing signals, which is called the intermediate frequency (IF) board.  The firmware on the FPGA is used to control the output of the DAC and to process the digitized signal from the ADC.  

The two outputs of the DAC make up the in-phase and quadrature (IQ) components of a complex signal used in a quadrature amplitude modulation strategy.  The DAC is used to output a frequency comb that is mixed with a local oscillator (LO) frequency to move the comb into the frequency range of the MKID resonator frequencies.  This is then sent through the MKID array.  Each frequency tone in the comb corresponds to one MKID resonator frequency.  The phase of the tone is set so the baseline phase of the same frequency read in the ADC is approximately zero.  The amplitude of the tone is set according to the optimal readout power of the resonator addressed relative to the other resonators. The complete frequency comb sent to the DAC is retrieved by the FPGA firmware from a look-up table (LUT) stored in memory on the ROACH board.

The DACs and ADCs are clocked at 512 MHz and the FPGA is clocked at 256 MHz.  The frequency comb output from each DAC can read out up to 256 resonators in 512 MHz of bandwidth, with 2 MHz spacing between tone frequencies.  The 2024 pixel array used in ARCONS is divided into two feedlines with 1012 pixels each.  Four readout boards are connected to each feedline through power combiners and power splitters.

Once the readout signal has passed through the MKIDs and has been mixed back down to baseband frequencies by the IF board and digitized by the ADC, the FPGA firmware divides the signal into 256 frequency bins (channels) through a ``channelization'' process so that the contribution from each probe tone in the comb can be treated separately.

The firmware then searches for indications of photon absorption in the phase of each channel's signal.  These are seen as a sudden decrease in the phase followed by an exponential decay back to the baseline, with the characteristic time determined by the quasiparticle recombination time.  A template of this phase pulse shape was made for each pixel and used to create a customized optimal filters for that pixel.  The firmware applies the optimal filter to a channel's signal, then searches for negative peaks in the filtered phase that pass below a set threshold value.  Each time the peak detection triggers, the firmware creates a 64-bit data packet containing the channel number (so that the photon can be identified with a particular pixel), a timestamp, and values specifying the height of the phase pulse peak and the baseline phase level before the pulse to help remove any 1/f phase fluctuations.  A dead time of 100 $\mu$s is added after a trigger to keep pixels with incorrectly set trigger thresholds from filling up the buffers with false pulses.  A more elegant way of dealing with incorrect thresholds will be developed for future versions of the readout firmware.

The ROACH boards are synchronized using the 1 PPS output of a Meinberg GPS170PEX GPS board and a Stanford Research FS725 Rubidium 10 MHz frequency standard, which is needed for the timestamps to be recorded with 2 $\mu s$ resolution.

The firmware outputs the data packets for the detected photons to two shared memory blocks on the ROACH.  A C program called PulseServer running on the powerPC located on each ROACH monitors the shared memory blocks.  When they are filled to a certain point, PulseServer sends the data over a TCP connection to the data acquisition and control computer.  A C program called PacketMaster, running on the data acquisition and control computer, receives the TCP data bundles from all eight ROACH boards.  It sifts through each photon packet and organizes them by pixel and second of arrival to write into an HDF\footnote{\url{http://www.hdfgroup.org/}} data file.  PacketMaster also creates images for the real-time display on the control computer.  The resulting HDF data file can later be processed with our software pipeline, detailed in Section~\ref{sec:pipeline}.

\subsection{Cryogenic Wiring}

ARCONS contains two independent microwave signal paths, which we refer to as feedline 1 and 2.  Each feedline reads out half of the array.  The signal path begins with a hermetic SMA bulkhead connector that brings the comb of frequencies into the dewar.  A laser welded 0.087" stainless steel semi-rigid coaxial cable brings the signals from room temperature to 4 K, with a heat sink clamp cooling only the outer conductor at the 50-70 K stage.  At 4 K there is a 30 dB attenuator, followed by an inner and outer DC block.  After the DC block a 0.062" NbTi/NbTi (outer conductor/center pin) coax brings the signals to the 1 K stage, where the coax is interrupted with a heatsunk SMA barrel connector.  A second NbTi coax brings the signals to 100 mK.  After passing through the array, the signal goes through another NbTi coax to 1 K, then a second to 4 K where it enters the input port of a HEMT amplifier~\citep{2009RScI...80d4702W}.  After the HEMT, another stainless coax brings the signals to a hermetic SMA bulkhead connector and outside the dewar.

\section{Data Processing Pipeline}
\label{sec:pipeline}

\begin{figure}[tbp]
\includegraphics[width=1.0\columnwidth]{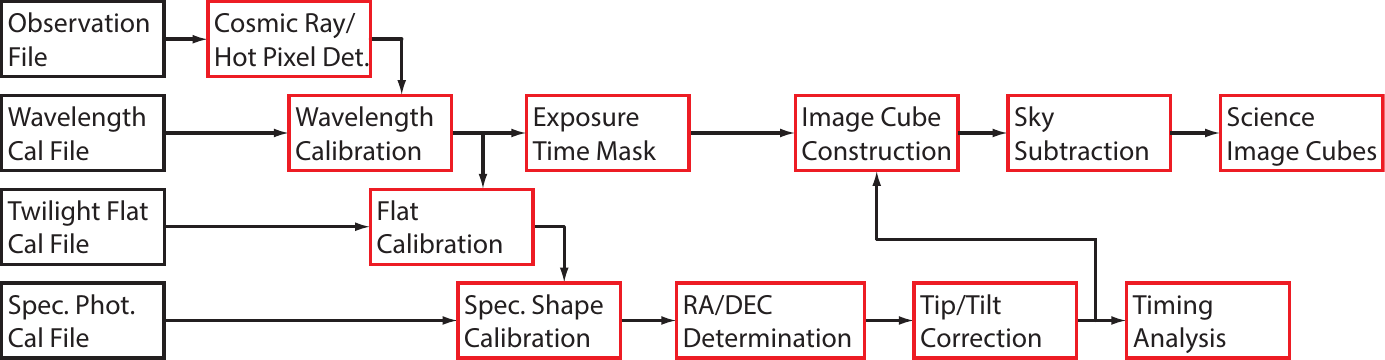}
\caption{\label{fig:pipeline}General outline of the ARCONS data
  reduction pipeline.}
\end{figure}

The ARCONS data reduction pipeline, though currently under development, is outlined here (see Figure \ref{fig:pipeline} for a schematic). The algorithms, written in Python, will be made publicly available once the pipeline has become stable and reliable.  Raw ``observation'' files consist of a set of tables, one for each pixel, with each row in the tables containing one second's worth of data packets, where each data packet represents a single photon event. These files form the raw input to the pipeline. The main steps in data reduction are as
follows.

\begin{description}
	
\item[Barycenter Photon Arrival Times]
The time of the start of each observation is stored in that observation HDF file's header.  The photon arrival timestamps recorded in the file are relative to this observation start time, so absolute arrival times at the location of the telescope are found by simply adding the recorded timestamps to the header time.  These timestamps can then be corrected to solar system barycentric times with the TEMPO2 pulsar timing package~\citep{hobbs:2006} using a custom plug-in that allows it to treat photons individually.   The barycentered timestamps can then be used to compare the timing between signals recorded from different observatory locations without the effects of the Earth's motion interfering.

\item[Cosmic Ray Cleaning] The photon arrival times across all pixels are binned into histograms with overlapping $(\sim10\,{\rm \mu s})$ bins, in order to search for multiple photon detection coincidences. Where an interval is found with an unusually high number of coincident photon detections, that interval of ${\sim}$100~$\mu$s is flagged as a cosmic-ray contaminated interval for the entire chip.  The cosmic ray event rate is low enough that the integration time lost to this procedure is negligible.

\item[Wavelength Calibration] Wavelength calibration files are created by uniformly exposing the detector array to the combined light of three lasers of known wavelength, as described in Section~\ref{sec:wavecal}. For each photon event, the measured phase excursion is assumed to be a monotonically increasing function of photon energy. A histogram of these phase shifts is created for each pixel. In a functional pixel the histogram shows a superposition of three approximately normal distributions corresponding to the three laser wavelengths. The locations of the maxima of these peaks are fit in order to find the phase shift corresponding to the three wavelengths. A second order polynomial fit to these three values, yields a smooth, continuous, analytic function, mapping phase shift to wavelength for each given pixel.

\item[Flatfield Calibration] Twilight sky exposures are used to normalize the response of each pixel with respect to each other.  The average twilight spectrum is computed by taking the median of the wavelength calibrated twilight spectrum for each pixel.  The ratio of this average spectrum to an individual pixel's wavelength calibrated spectrum then gives a weighting as a function of wavelength for that pixel.  This weighting is used to normalize the response of different pixels when combining data from multiple pixels, effecting the traditional CCD ``flatfielding'' but at every wavelength.

\item[Spectral Shape Calibration] Once the pixel responses are normalized relative to one other, a final absolute calibration of the overall spectral response is performed by comparing the spectrum of an observed reference target against its known spectrum. This provides a single additional whole-array wavelength weighting function. The combination of this absolute weighting and the per-pixel relative flatfielding weights corresponds to the overall QE response of each pixel as a function of wavelength. Applied together, the weights yield a fully calibrated spectrum for each pixel.

\item[Exposure Time Masking]  High levels of photon illumination induce a number of the pixels in the array to go ``hot'' from time to time, registering false counts and reporting artificially high count rates.  This is likely due to an interaction of the MKID with free electrons in the silicon wafer, which should be eliminated in new MKID designs.  This pixel behavior is detected by comparing the flux distribution in the neighborhood of each pixel to that of the expected PSF (similar to the algorithm used in the XMM-Newton pipeline).\footnote{XMM-Newton Science Analysis System,  Users Guide to the XMM-Newton Science Analysis System”, Issue 9.0, 2012 (ESA: XMM- Newton SOC). See also \url{http://xmm.esac.esa.int/sas/}.} In each one-second time step, the ratio of the counts in each pixel to the median counts in a small surrounding box (typically $5\times5$ pixels) is compared to the same ratio for a 2-dimensional Gaussian function with FWHM matching that of the expected PSF (specified manually at the time of the data reduction). If, after accounting for photon shot noise errors, the measured ratio is significantly greater (typically $>3\sigma$) then the source of the counts cannot be astrophysical, and the pixel is flagged as hot for the given timestep. Some pixels are also known to show anomalously \textit{low} count rates (``cold'' pixels), or no counts at all (``dead'' pixels).  In a similar way, therefore, if the count rate for a pixel is significantly lower than the median for the surrounding box, the pixel can be flagged as ``cold''. These intervals are then concatenated and tracked to maintain consistency in the total  effective exposure time attributable to each pixel. Since the number of bad pixels is relatively small compared to the total number of pixels, and the time intervals over which such unwanted behavior occur are small, the time masks are most efficiently stored as per-pixel lists of bad intervals, also in HDF file format. This process is applicable to any array observation, including the laser wavelength calibrations and the twilight flats.

\item[RA/Dec Determination] An astrometric solution is found for the data at regular intervals (typically 30\,s). The centroid location of the target source in pixel coordinates is found using the Python language ``PyGuide'' centroiding algorithm,\footnote{See \url{http://www.astro.washington.edu/users/rowen\\/PyGuide/Manual.html}} which calculates the point of minimum radial asymmetry for a source in a time integrated image. The known sky coordinates of the target can then be assigned to that pixel location. Since the instrument location at the Coud\'{e} focus is fixed with respect to the telescope motion, the field of view also rotates at the sidereal rate. The hour angle of the target therefore relates directly to the orientation of the field. In combination with a pre-calibrated fixed rotational offset for the instrument, a right ascension and declination can be determined for every pixel at each time step.  In addition, atmospheric refraction spreads light of different frequencies onto different parts of the array.  This phenomenon is well understood~\citep{Auer:2000cr} and can be easily removed, allowing coordinates to be assigned for every photon.  

\item[Tip/Tilt Correction] For sources with sufficient count rates it should be possible to perform the above centroiding procedure on very short timescales, comparable with the timescale of atmospheric seeing variation ($\sim 10\mathrm{\,ms}$). Alternatively, for non-point-source objects, image cross-correlation can be performed to determine offset corrections to the data on the same timescales. Hence it should be possible in principle to apply \textit{ex post facto} first-order tip-tilt seeing corrections, purely in software, to improve the final recovered PSF.

\end{description}

After these steps have been performed, processed photon lists will form the output data product: calibrated lists of photon events, with each photon having an absolute timestamp, RA and Dec location, wavelength, and any associated flags for each photon, along with associated time-mask lists. 

Further analysis depends on the science goals and the desired end product.  For imaging and spectroscopic analysis, image cubes are constructed by creating images integrated over different wavelength ranges and over the desired time intervals. Background subtraction is performed as in traditional CCD data reduction (for example, by subtracting a constant or slowly varying two-dimensional polynomial fit to the image after masking bright sources). Spectra are then obtained by integrating over the desired pixels in the images.

Alternatively, science involving timing analysis (e.g., pulsar and eclipsing binary observations) requires a non-discrete approach in the time-axis. We integrate first over selected pixels or a selected area in sky-coordinates, and maintain the individual photon timing information, avoiding binning into larger time intervals. In the case of pulsar or eclipsing binary observations, for example, this allows for folding of the data on the known period.

\section{On-Sky Performance}
\label{sec:perf}

Even with the current engineering-grade MKID arrays, the extra mirror bounces and path length of the Coud\'{e} focus, and non-optimal optical design and anti-reflection coatings, ARCONS is one of the most powerful tools for narrow field of view astrophysics ever developed.  It has now been proven in the field over the course of 24 observing nights at 5-m class telescopes.  In this section we explore in detail the performance of the instrument.

\subsection{Array Yield}
\label{sec:yield}

\begin{figure}
\begin{center}
\includegraphics[width=0.8\columnwidth]{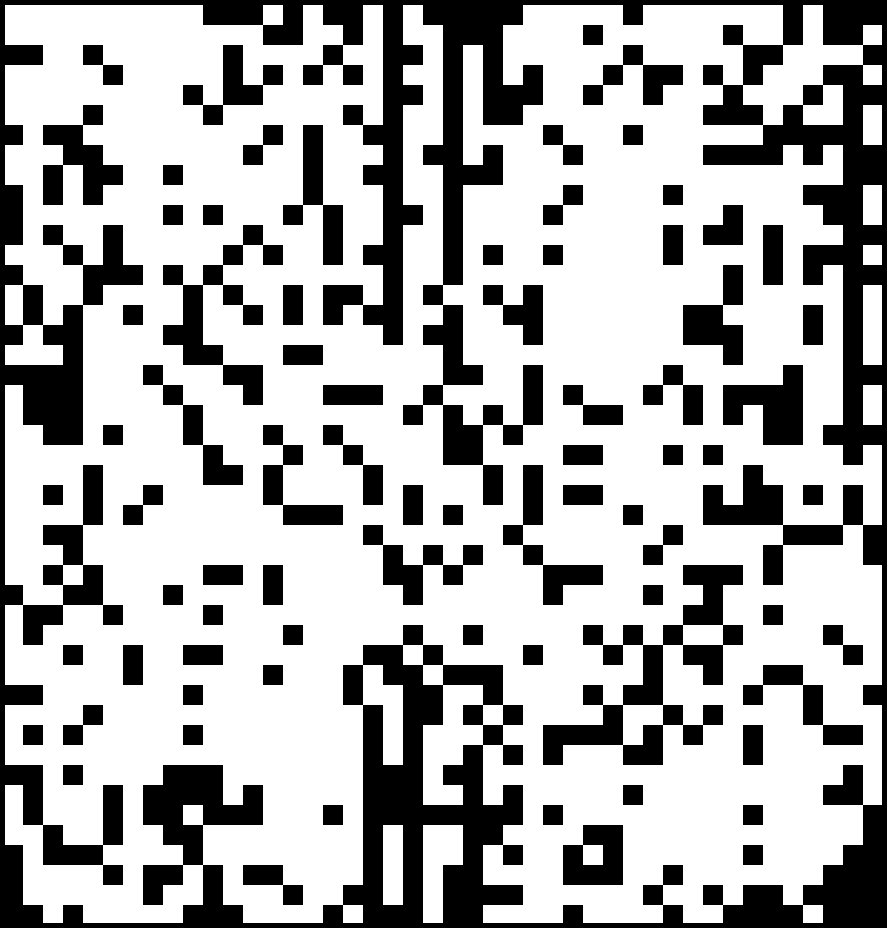}
\end{center}
\caption{The results of beam mapping the ARCONS array.  Pixels with good locations are shown in white.  Vignetting is apparent at the bottom right side of the array.  The overall pixel yield in this engineering-grade array is ${\sim}70$\%.}
\label{fig:yield}
\end{figure}

Typically over 90\% of the resonators show up in frequency sweeps, but due to the variations in the TiN gap there are a significant number of collisions, reducing usable pixels, as shown in Figure~\ref{fig:yield}.  After cutting out collisions and pixels with especially high or low quality factor, usually about 70\% of our pixels are usable. A more uniform film, such as the ones made at NIST with Ti/TiN multilayers~\citep{Vissers:2012wu} or with atomic layer deposition (ALD), could significantly reduce the number of collisions, dramatically improving yield.

\subsection{Spectral Resolution}
\label{sec:res}

\begin{figure}
\begin{center}
\includegraphics[width=1.0\columnwidth]{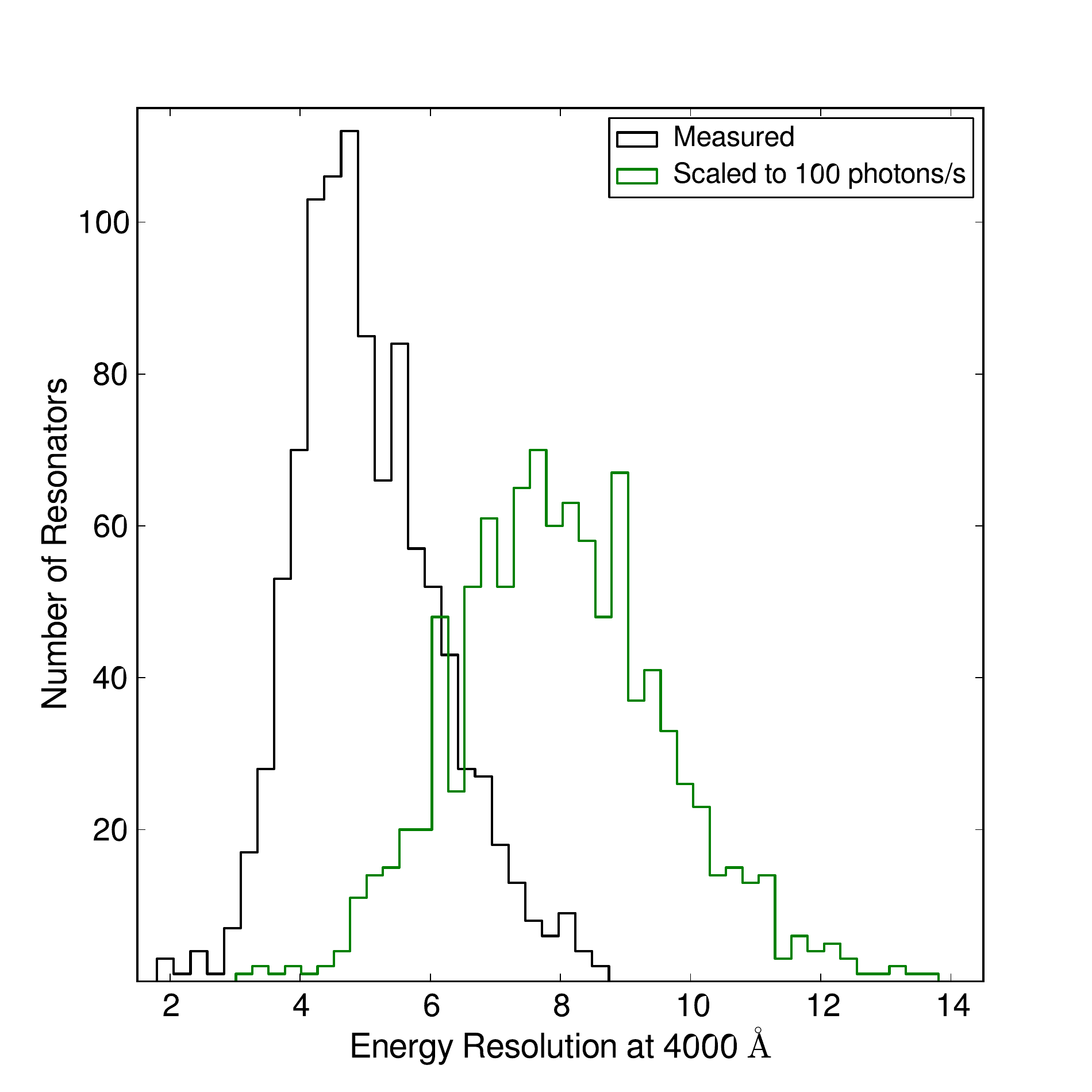}
\end{center}
\caption{A histogram of spectral resolution of one feedline from the array.  The black line shows the energy resolution from calibration files taken at Palomar with a high count rate of ${\sim}1000$ cts/sec.  The green curve shows the projected energy resolution we expect to recover at the nominal sky count rate of 100 cts/sec by scaling the black data by the expected degradation in spectral resolution with count rate, the red curve in Figure~\ref{fig:lin}.  The other feedline is very similar.}
\label{fig:res}
\end{figure}

The typical spectral resolution for the MKIDs used in ARCONS is R${\sim}$10 at 4000 \AA, declining linearly with increasing wavelength as expected.  The median spectral resolution is degraded to R${\sim}$8 with our two channel analog electronics (see Figure~\ref{fig:res}) with the full ARCONS digital readout due to limitations in the number of coefficients (taps) in the programmable optimal filter in our digital electronics.  This issue is being addressed by upgrading our electronics from ROACH to ROACH2, which will allow for significantly more taps.

The achieved spectral resolution is drastically below the theoretical limit for a 100 mK operating temperature of R=100 at 4000 \AA~\citep{Mazin:2012kl}.  We believe the current MKIDs are limited by positional effects within the MKID inductor due to local variations in the TiN superconducting gap.  Fixing this should roughly double the spectral resolution.  After this issue is addressed, lower noise microwave amplifiers~\citep{HoEom:2012kq} and new MKID designs should continue to boost the resolution towards the theoretical limit.

As shown in Figure~\ref{fig:lin} as a red line (labeled on the right y-axis), there is also a pronounced degradation of the spectral resolution with increasing count rate.  This is caused by two effects.  First, the baseline value that the firmware tracks to reduce 1/f noise gets noisy from averaging many counts.  Second, pulses begin overlapping, degrading resolution.  Improvements in the firmware baseline determination algorithm should improve this degradation significantly.  At Palomar, the ARCONS sky background rate is 50--100 photons/sec, so only objects significantly brighter than sky will suffer from severe degradation of the spectral resolution.

\subsection{Sensitivity}
\label{sec:sens}

\begin{figure}
\begin{center}
\includegraphics[width=1.0\columnwidth]{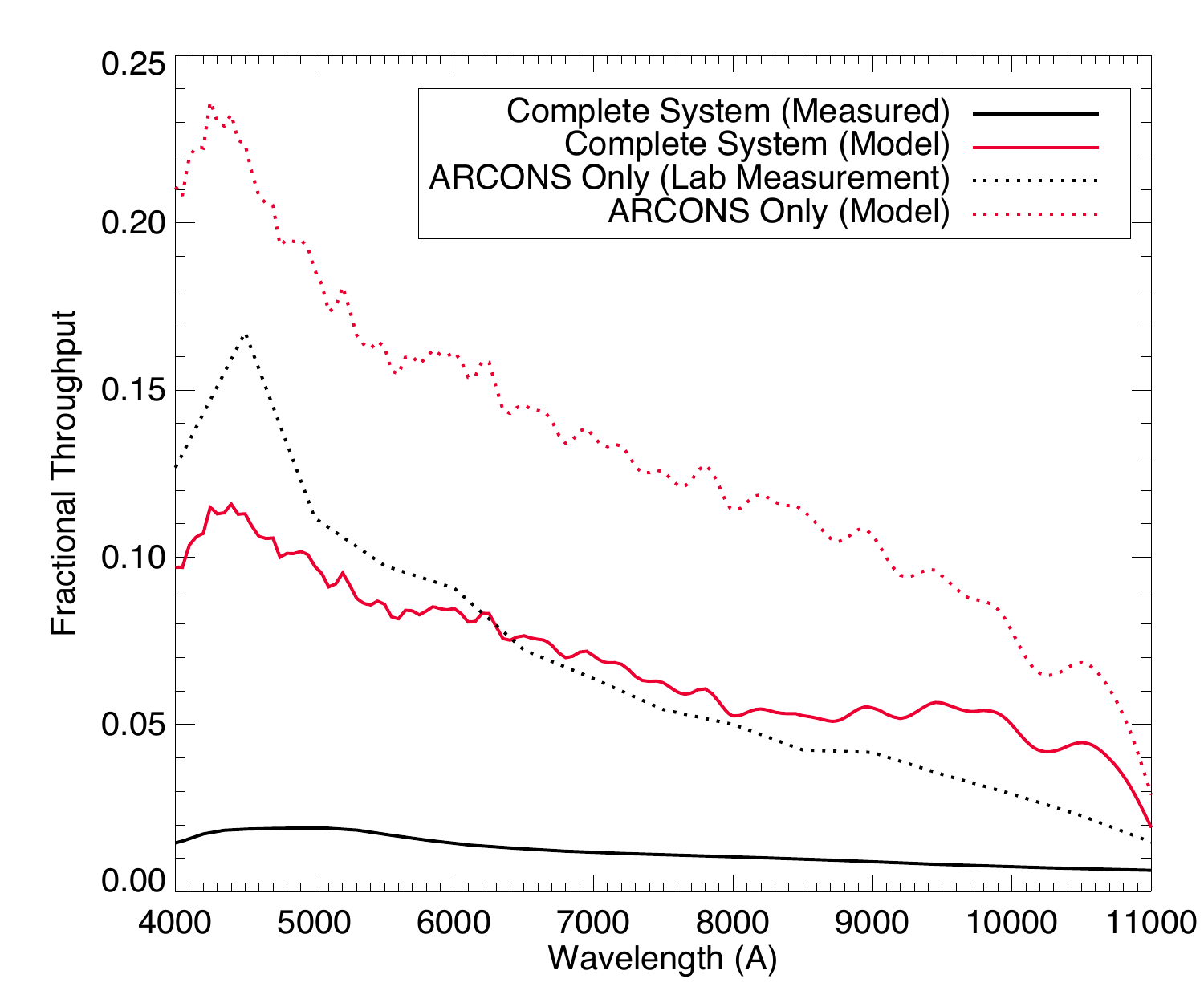}
\end{center}
\caption{The system throughput of ARCONS.}
\label{fig:sens}
\end{figure}

The MKIDs in ARCONS absorb light directly in the TiN film that comprises the inductor.  This TiN has an intrinsic absorption of roughly 70\% at 4000 \AA, and 30\% at 1~\micron.  The total system throughput is reduced by the cumulative effect of all the intervening optical materials and surfaces; the atmosphere, the telescope M1--M4 mirrors (the Coud\'{e} M3 at Palomar was degraded due to $>10$ years since realuminization), the collimating lens, the Asahi Supercold Filter at 4 K, the 4 optical elements at 4K, the 100 mK glass with ITO filter, and the microlens fill factor and alignment with the inductors.  A model of this entire system, based on a simple atmospheric model, clean Al reflectivity (normalized to 86\% at 670 nm), manufacturer's specs for glasses, AR coatings, and the Supercold Filter, and measurements of the absorptivity of raw TiN films, is shown as a red solid line in Figure~\ref{fig:sens}. The efficiency of just ARCONS, neglecting the atmosphere and telescope, is shown as the red dashed line.

The theoretical model of ARCONS can be directly compared with laboratory measurements of ARCONS's QE. These tests were done with a QE testbed consisting of a monochromater and light source, integrating sphere, and a rotatable fold mirror to direct the beam to a calibrated photodiode or to the ARCONS optical input.  The results of these lab tests are shown as a black dashed line in Figure~\ref{fig:sens}.  The roughly factor of two discrepancy between the theoretical model and the laboratory measurements is primarily due to a known misalignment of the microlens height above the focal plane, increasing the spot size on the focal plane so it is larger than the inductor, especially at red wavelengths.  

The actual system throughput is calculated by doing photometry on a standard star and comparing our results to the object's known spectrum, shown as the solid black line in Figure~\ref{fig:sens}.  This curve is about a factor of two below the laboratory QE measurements modified by the expected atmospheric transmission and telescope throughout.  The cause of this discrepancy is unknown, but could include the unknown reflectivity of the Coud\'{e} M3 flat or angular misalignment of ARCONS with respect to the telescope input. 

The MKIDs used is ARCONS have inductors with a preferred orientation of the wires, which could potentially lead to a quantum efficiency that depends on polarization angle.  Lab measurements were performed on ARCONS to check whether the MKID arrays exhibited any response to linearly polarized light at different angles.  No polarization response was seen to the 5\% level.

By going to Cassegrain focus, improving our anti-reflection coatings, aligning the microlens well, and simplifying our optics we should be able to boost our total system throughput in future MKID cameras by a factor of 10.  Using MKIDs with black absorbers could eventually yield instruments with total system efficiencies greater than 60\%.

\subsection{Linearity}
\label{sec:lin}

\begin{figure}
\begin{center}
\includegraphics[width=1.0\columnwidth]{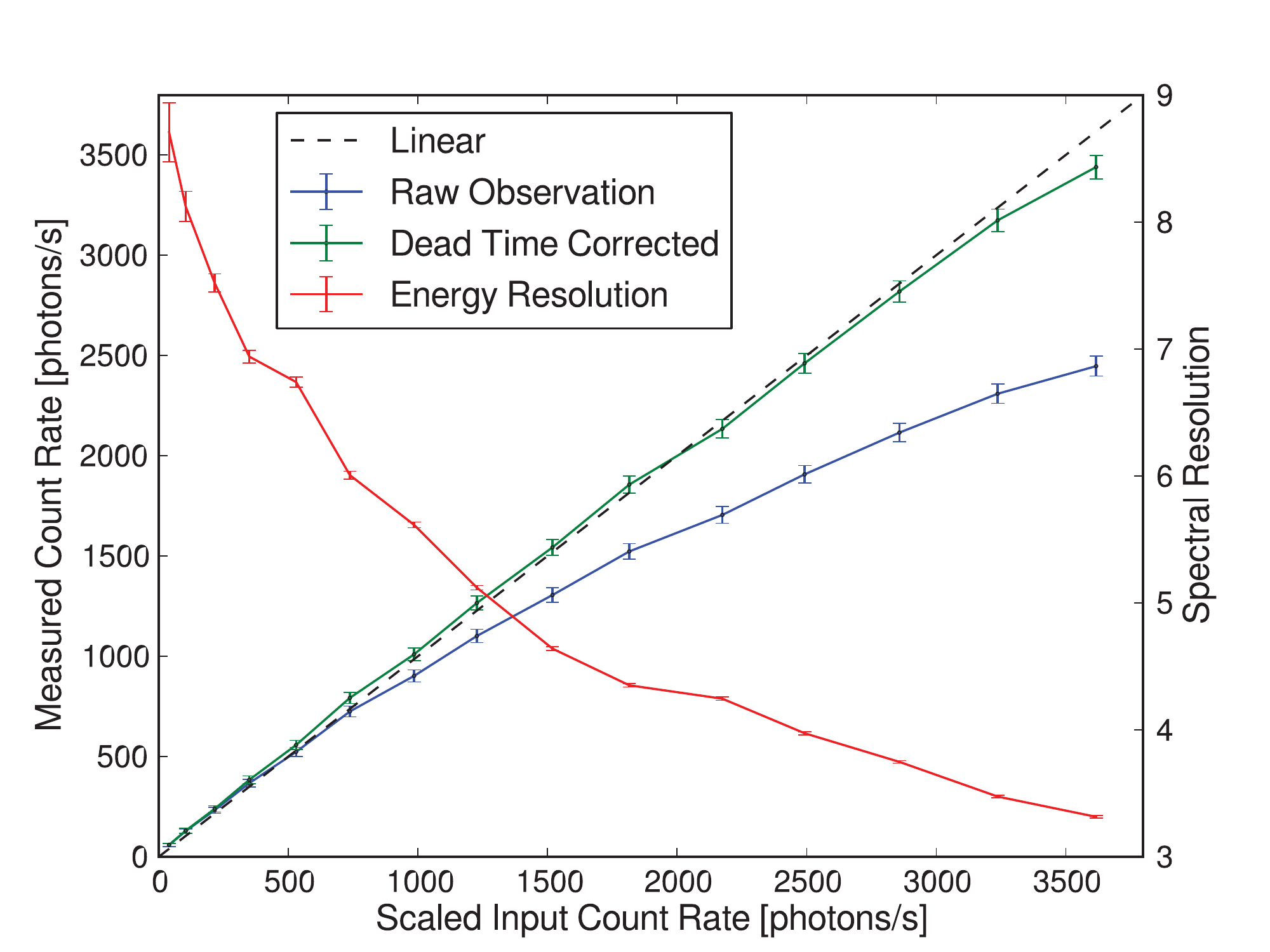}
\end{center}
\caption{The linearity and spectral resolution vs. count rate of the MKID array in ARCONS measured in the lab.  The measured count rate is the count rate actually recorded by the MKID, and the scaled input count rate is the count rate we expect to measure based on the measured flux in a calibrated photodiode.}
\label{fig:lin}
\end{figure}

The 100 $\mu$s dead time associated with triggering (Section~\ref{sec:readout}) causes some pulses to be missed at high count rates, as shown in Figure~\ref{fig:lin}.  This effect can be corrected for analytically by calculating the number of photons that are expected to arrive within a dead time interval and hence not be counted,  $R_{c}=R_m/(1-R_m d)$, where $R_c$ is the corrected count rate, $R_m$ is the measured count rate, and $d$ is the dead time, as shown in as a green line in Figure~\ref{fig:lin}.  The slightly longer effective dead time of 120 $\mu$s used in Figure~\ref{fig:lin} results from the interaction of the firmware dead time and the pulse optimal filters.   The firmware of ARCONS is currently being rewritten to remove the dead time and better handle closely spaced photons, which should significantly improve the linearity and spectral resolution at high count rates.

\subsection{Imaging and Spectrophotometry}
\label{sec:imaging}

\begin{figure}
\begin{center}
\includegraphics[width=1.0\columnwidth]{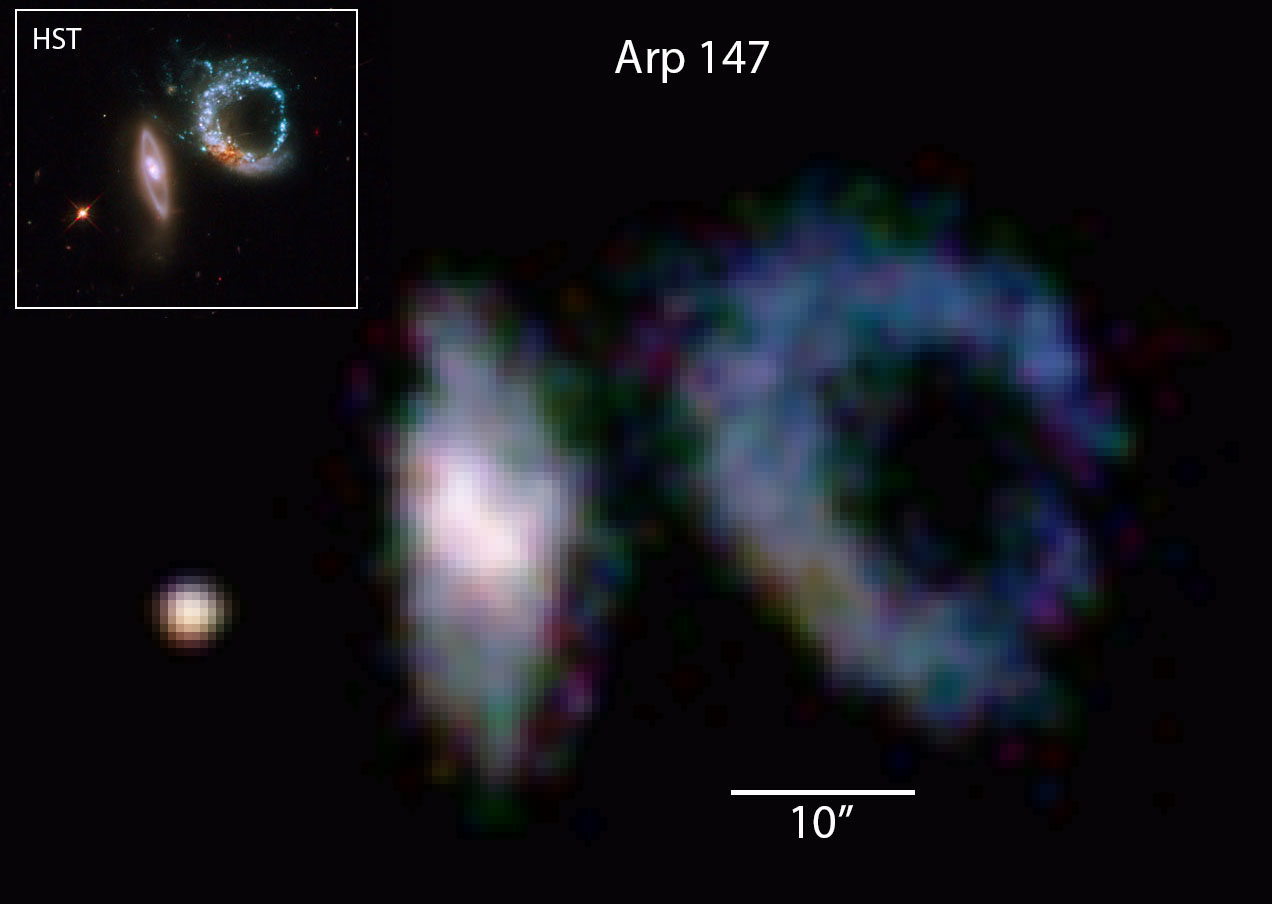}
\end{center}
\caption{A mosaic of the interacting galaxies Arp 147 made with ARCONS on the Palomar 200".  The data consists of 36 pointing of 1 minute each.  To make this image, the data was cleaned of cosmic rays and hot pixels, wavelength calibrated, and flat fielded.  This processed data for each of the 36 pointings was offset in RA and Dec by the amount requested by our mosaicing control software and combined.  No field derotation was performed.  The false colors were made by breaking the ARCONS data into three wavelength bands and setting each band to correspond to to appropriate RGB value.  The image was convolved with a Gaussian to reduce noise below the spatial scale of the seeing, and the image was adjusted in Photoshop to give an attractive color palette and remove some minor sky noise artifacts.  The inset shows a processed HST image of Arp 147.}
\label{fig:arp}
\end{figure}

Imaging and spectroscopic performance of ARCONS is consistent with expectations from the beam map and spectral resolution data.  The plate scale was measured to be 0.435"/pixel, matching well with our Zemax design.  During the run we experienced one night with seeing measured at $\sim$1.0" by the Palomar seeing monitor, and fitting a PSF from this night shows that we recover this seeing value, indicating that our optics are not limiting our imaging performance.

\begin{figure}
\begin{center}
\includegraphics[width=1.0\columnwidth]{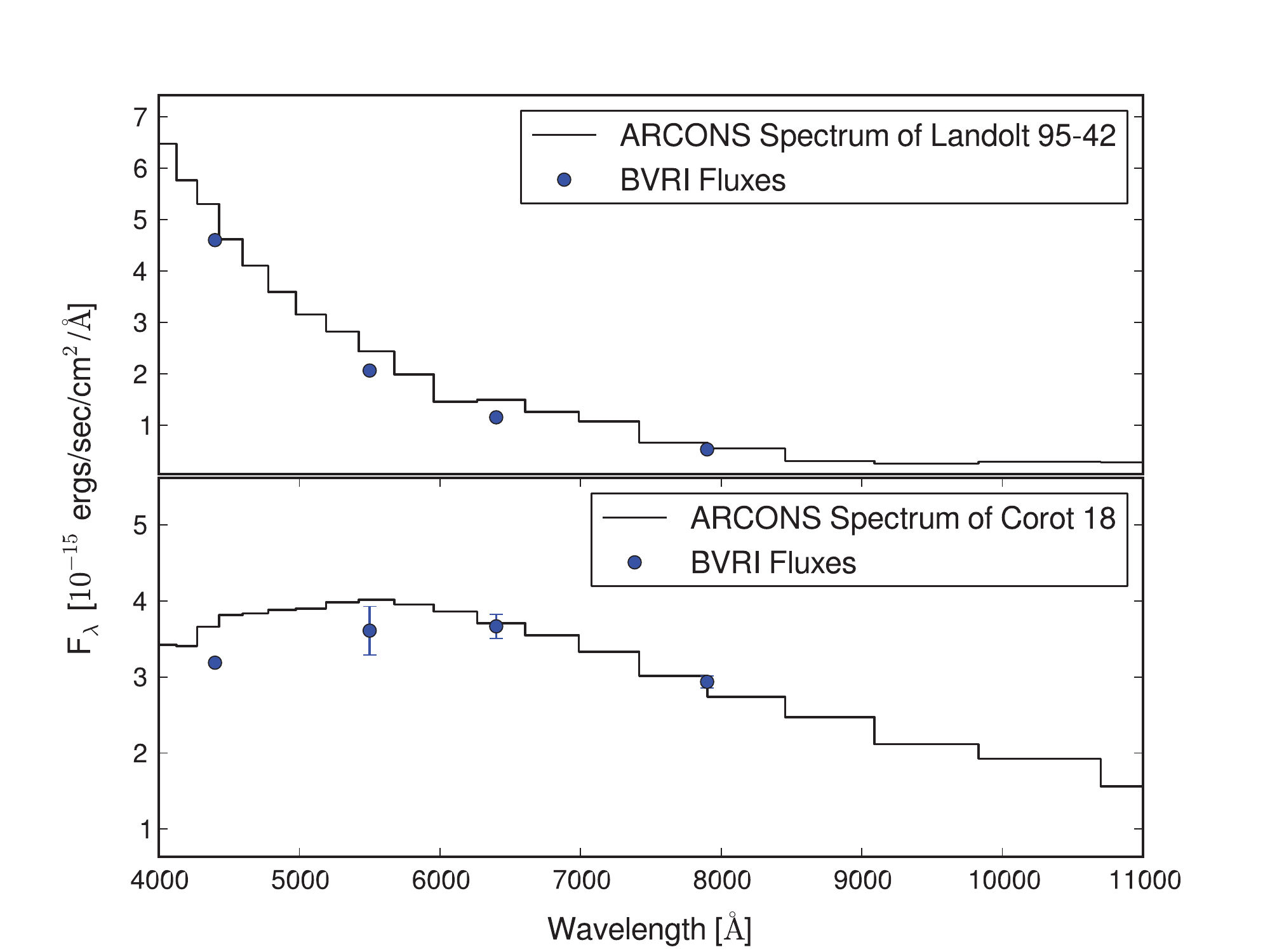}
\end{center}
\caption{Spectra of Landolt 94-42~\citep{1992AJ....104..340L} on top, and the G9V star Corot-18 on the bottom, taken with ARCONS on the Palomar 200".  These spectra are flux calibrated.  Previous BVRI photometry is shown as blue points.}
\label{fig:spectra}
\end{figure}

Extremely high quality imaging and spectra will require more of our software pipeline to be in place, but preliminary analyses are presented in Figures~\ref{fig:arp} and~\ref{fig:spectra}.  These images and spectra begin to show the capabilities of this unique instrument.

\subsection{Timing}
\label{sec:timing}

\begin{figure}
\begin{center}
\includegraphics[width=1.0\columnwidth]{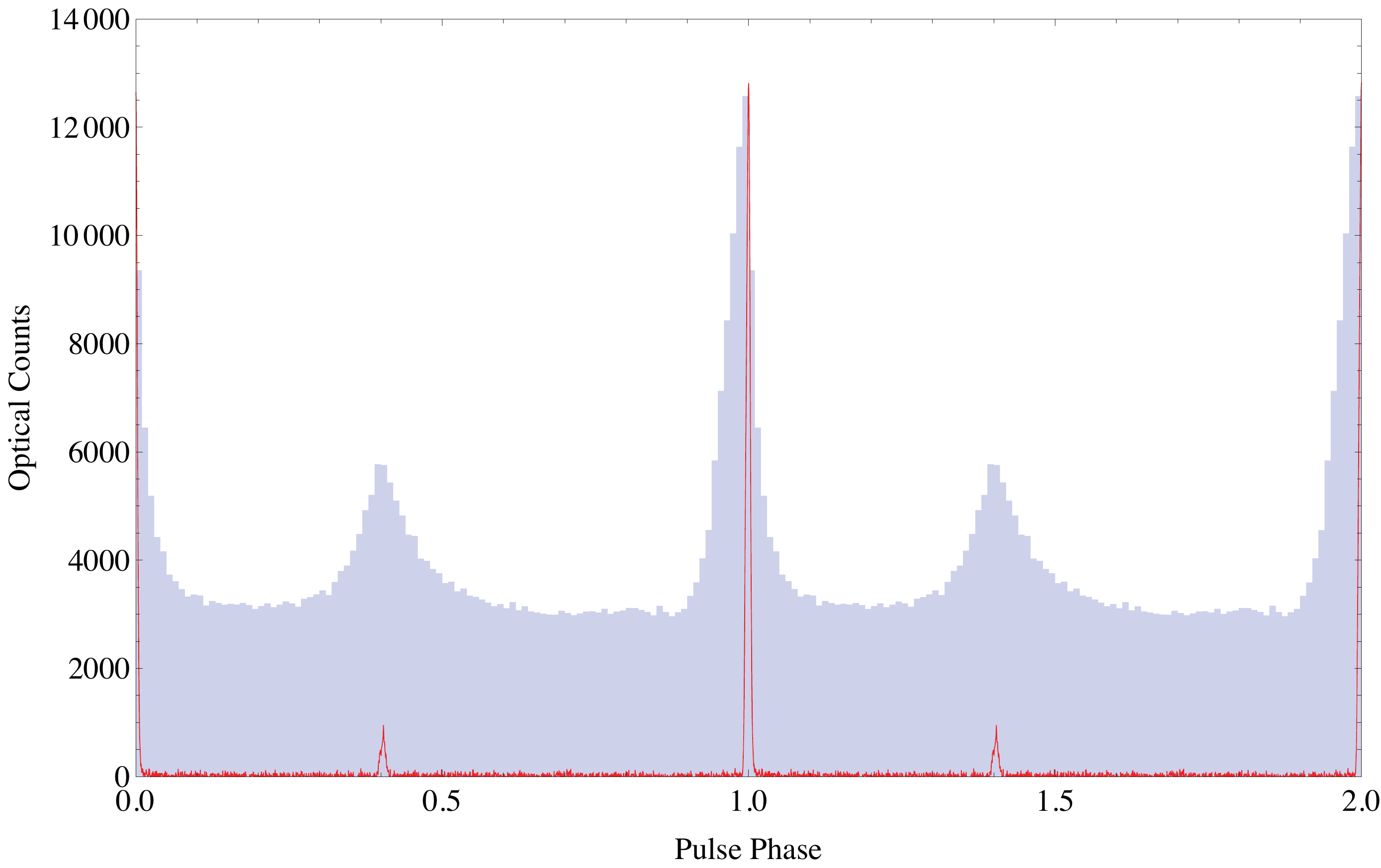}
\end{center}
\caption{The alignment of the radio (red) and optical (blue) pulses from the Crab Pulsar verify the absolute timing accuracy of ARCONS.}
\label{fig:crab}
\end{figure}

ARCONS timing performance was measured by simultaneously observing the Crab Pulsar with ARCONS and the Green Bank Telescope (GBT).  The data shown in Figure~\ref{fig:crab} shows that ARCONS sees an optical pulse leading the radio pulse by about 150~$\mu$s, which is consistent with previous observations~\citep{Shearer:2003uc}.  

Measurement of the arrival time of a single photon should be accurate ${\sim}2~\mu$s.  Currently, there is a delay of 43~$\mu$s from the time a photon is absorbed by the MKID until the photon packet is ready to be sent over the network, opening up the possibility of extremely fast effective frame rates for certain applications, like wavefront sensing and speckle control in a coronagraph.

\section{Conclusions}
\label{sec:conc}

ARCONS is a unique, read-noise free photon counting IFS.  With a 2024 MKID pixel array covering a 20"${\times}$20" field of view, ARCONS is the world's largest (and only active) LTD-based instrument in the optical through near-IR.  It is uniquely powerful for observations of rapidly variable sources. 

The ARCONS instrument has performed astronomical observations, and the first science papers using ARCONS data are currently being written. ARCONS and its successors will continue to improve in pixel count and yield, spectral resolution, system throughput, and detector quantum efficiency. The MKID technology it uses is extremely scalable, allowing arrays approaching megapixels within a decade.  MKID-based instruments like ARCONS will start to become a standard part of the UVOIR observer's toolkit in the years to come.
 
\section*{Acknowledgments}

The MKID detectors used in this work were developed under NASA grant NNX11AD55G.  S.R. Meeker was supported by a NASA Office of the Chief Technologist’s Space Technology Research Fellowship, NASA grant NNX11AN29H.  This work was partially supported by the Keck Institute for Space Studies.  Fermilab is operated by Fermi Research Alliance, LLC under Contract No. De-AC02-07CH11359 with the United States Department of Energy.  The authors would also like to thank Shri Kulkarni, Director of the Caltech Optical Observatories, and Jason Prochaska, Associate Director of Lick Observatory, for facilitating this project, as well as the excellent staffs of the Palomar and Lick Observatories for their assistance in getting ARCONS working.  Jennifer Milburn's help with the guide camera software was invaluable.  This project also greatly benefitted from the support of Mike Werner, Paul Goldsmith, and Jonas Zmuidzinas at JPL.

\bibliographystyle{apj}

\end{document}